\documentclass[aps,prb,twocolumn,floatfix,footinbib,showpacs,superscriptaddress]{revtex4}
\usepackage[english]{babel}
\usepackage[utf8]{inputenc}
\usepackage{graphicx}
\usepackage{fancyhdr}
\usepackage[active]{srcltx}
\usepackage{setspace}
\usepackage{float}
\usepackage{amsmath}
\usepackage{amsfonts}
\usepackage{natbib}
\usepackage{color}
\begin{document}

\title{Toward high-frequency operation of spin lasers}

\author{Paulo E. Faria Junior}
\affiliation{Department of Physics, University at Buffalo, State University of New York, Buffalo, New York 14260, USA}
\affiliation{Instituto de F\'isica de S\~ao Carlos, Universidade de S\~ao Paulo, 13566-590 S\~ao Carlos, S\~ao Paulo, Brazil}

\author{Gaofeng Xu}
\affiliation{Department of Physics, University at Buffalo, State University of New York, Buffalo, New York 14260, USA}

\author{Jeongsu Lee}
\affiliation{Department of Physics, University at Buffalo, State University of New York, Buffalo, New York 14260, USA}
\affiliation{Institute for Theoretical Physics, University of Regensburg, 93040 Regensburg, Germany}

\author{Nils C. Gerhardt}
\affiliation{Department of Physics, University at Buffalo, State University of New York, Buffalo, New York 14260, USA}
\affiliation{Photonics and Terahertz Technology, Ruhr-University Bochum, D-44780 Bochum, Germany}

\author{Guilherme M. Sipahi}
\affiliation{Department of Physics, University at Buffalo, State University of New York, Buffalo, New York 14260, USA}
\affiliation{Instituto de F\'isica de S\~ao Carlos, Universidade de S\~ao Paulo, 13566-590 S\~ao Carlos, S\~ao Paulo, Brazil}

\author{Igor \v{Z}uti\'{c}}
\affiliation{Department of Physics, University at Buffalo, State University of New York, Buffalo, New York 14260, USA}

%===============================================================================

%put NEW figs, phonon laser, other refs

\begin{abstract}

Injecting spin-polarized carriers into semiconductor lasers provides important 
opportunities to extend what is known about spintronic devices, as well as to 
overcome many limitations of conventional (spin-unpolarized) lasers. By developing 
a microscopic model of spin-dependent optical gain derived from an accurate electronic 
structure in a quantum well-based laser, we study how its operation properties 
can be modified by spin-polarized carriers, carrier density, and resonant cavity 
design. We reveal that by applying a uniaxial strain, it is possible to attain a 
large birefringence. While such birefringence is viewed as detrimental in conventional 
lasers, it could enable fast polarization oscillations of the emitted light in 
spin lasers which can be exploited for optical communication and high-performance 
interconnects. The resulting oscillation frequency ($>200$ GHz) would significantly 
exceed the frequency range possible in conventional lasers.

\end{abstract}

\pacs{42.55.Px, 78.45.+h, 78.67.De, 78.67.Hc}

\maketitle

%===============================================================================

\section{Introduction}

Both spin lasers and their conventional (spin-unpolarized) counterparts share 
three main elements: (i) the active (gain) region, responsible for optical 
amplification and stimulated emission, (ii) the resonant cavity, and (iii) the 
pump, which injects (optically or electrically) energy/carriers. The main distinction 
of spin lasers is a net carrier spin polarization (spin imbalance) in the active 
region, which, in turn, can lead to crucial changes in their operation, as compared 
to their conventional counterparts. This spin imbalance is responsible for a circularly 
polarized emitted light, a result of the conservation of the total angular momentum 
during electron-hole recombination.\cite{Meier:1984} 

The experimental realization of spin lasers\cite{Hallstein1997:PRB, Ando1998:APL, 
Rudolph2003:APL,Holub2007:PRL,Hovel2008:APL,Basu2008:APL,Basu2009:PRL, Fujino2009:APL,
Saha2010:PRB, Jahme2010:APL,Gerhardt2011:APL, Iba2011:APL,Frougier2013:APL, Frougier2015:OE,
Hopfner2014:APL, Cheng2014:NN, Alharthi2015:APL,Hsu2015:PRB} 
presents two important opportunities.
The lasers provide a path to practical room-temperature spintronic devices with different 
operating principles, not limited to magnetoresistive effects, which have enabled 
tremendous advances in magnetically stored information.\cite{Zutic2004:RMP, 
Fabian2007:APS, Tsymbal:2011,Maekawa:2002,DasSarma2000:SM} %I
This requires revisiting the common understanding of material parameters 
for desirable operation,\cite{Lee2014:APL} as well as 
a departure from more widely studied spintronic devices, where only one type 
of carrier (electrons) plays an active role. In contrast, since semiconductor lasers 
are bipolar devices, a simultaneous description of electrons and holes is crucial. 

On the other hand, the interest in spin lasers is not limited to spintronics, as 
they may extend the limits of what is feasible with conventional semiconductor 
lasers. It was experimentally demonstrated that injecting spin-polarized carriers 
already leads to noticeable differences in the steady-state operation.\cite{Rudolph2003:APL,
Holub2007:PRL, Hovel2008:APL} The onset of lasing is attained for a smaller injection, 
lasing threshold reduction, while the optical gain differs for different polarizations 
of light,  %I
leading to gain asymmetry, 
also referred to as gain anisotropy.\cite{Holub2007:PRL, Hovel2008:APL, 
Basu2009:PRL} In the stimulated emission, even a small carrier polarization in 
the active region can be greatly amplified and lead to the emission of completely 
circularly polarized emitted light, an example of a very efficient spin filtering.\cite{Iba2011:APL} 

\begin{figure}[h!]
\begin{center} 
\includegraphics{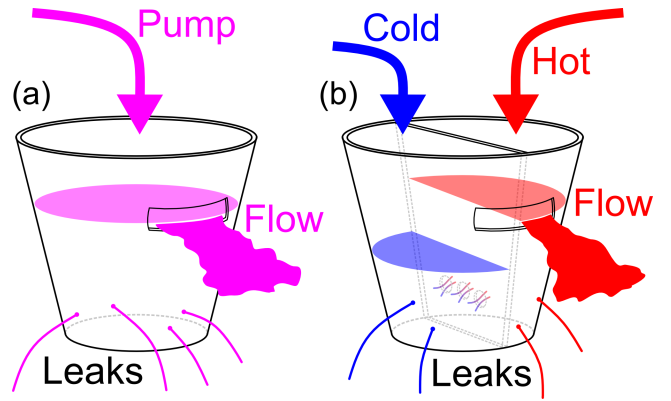}
\caption{(Color online) Bucket model for (a) a conventional laser and (b) a spin laser.\cite{Lee2012:PRB} 
Water added to the bucket represents the carriers, and the water coming out represents the 
emitted light. Small leaks depict spontaneous emission, and overflowing water reaching 
the large opening corresponds to the lasing threshold. In (b) the two halves represent 
two spin populations (hot and cold water in the analogy), and they are filled separately. 
The partition between them is not perfect: spin relaxation can cause the two 
populations to mix. The color code indicates conservation of angular momentum; 
an unpolarized pumping (violet) is an equal mixture of two polarized 
contributions (red and blue).}
\label{fig:bucket}
\end{center}
\end{figure}

An intuitive picture for a spin laser is provided by a bucket model in 
Fig.~\ref{fig:bucket}.\cite{Lee2012:PRB,Zutic2014:NN} The uneven water levels represent the 
spin imbalance in the laser, which implies the following: (i) Lasing threshold reduction - in a 
partitioned bucket, less water needs to be pumped for it to overfill. There are 
also two thresholds (for cold and hot water).\cite{Gothgen2008:APL} (ii) Gain 
%I nisotropy 
asymmetry - an unequal amount of hot and cold water comes out. A small spin 
imbalance of pumped carriers can (the two water levels slightly above and below 
the opening, respectively) result in a complete imbalance in the polarization 
of the emitted light (here only hot water gushes out) and, consequently, spin-filtering. 
These effects are attained at room temperature with either optical or electrical 
injection. The latter experimental demonstration\cite{Cheng2014:NN} is a %I n important 
breakthrough towards practical use of spin lasers. 

Perhaps the most promising opportunity to overcome the limitations of conventional 
lasers lies in the dynamic operation of spin lasers, predicted to provide enhanced 
modulation bandwidth, improved switching properties, and reduced parasitic frequency 
modulation, i. e., chirp.\cite{Lee2014:APL,Lee2012:PRB,Lee2010:APL,Boeris2012:APL} Moreover, 
experiments have confirmed that in a given device a characteristic frequency of 
polarization oscillations of the emitted light can significantly exceed the 
corresponding frequency of the intensity oscillations.\cite{Jahme2010:APL, 
Gerhardt2011:APL, Hopfner2014:APL} This behavior was attributed to birefringence 
- an anisotropy of the index of refraction, %I keep the anisotropy here
considered detrimental in conventional 
lasers.\cite{Chuang:2009}

What should we then require to attain high-frequency operation in spin lasers? 
Can we provide guidance for the design of an active region and a choice of the 
resonant cavity? Unfortunately, to address similar questions, we cannot simply 
rely on the widely used rate-equation description of 
spin lasers,\cite{Rudolph2003:APL, Holub2007:PRL,Lee2012:PRB,SanMiguel1995:PRA,Gahl1999:IEEEJQE}  
but instead we need to formulate a microscopic description. The crucial consideration 
is detailed knowledge of the spectral (energy-resolved) optical gain obtained 
from an accurate description of the electronic structure in the active region, 
already important to elucidate a steady-state operation of a spin laser.

\begin{figure}[h]
\begin{center}
\includegraphics{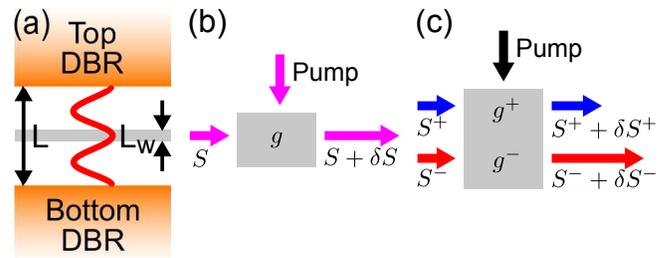}
\caption{(a) (Color online) Geometry of a vertical cavity surface-emitting laser. 
The resonant cavity of length $L$ is formed between the two mirrors made of distributed 
Bragg reflectors (DBRs). The shaded region represents the active (gain) region of length 
$L_W$. The profile of a longitudinal optical mode is sketched. Schematic of the 
optical gain, $g$, in the active region for a conventional laser (b) and a spin laser (c). 
With external pumping/injection, a photon density $S$ increases by $\delta S$ 
as it passes across the gain region.\cite{note_expgain} 
 In the spin laser, this increase depends on 
the positive ($+$)/negative($-$) helicity of the light, $S=S^+ +S^-$.}
\label{fig:VCSEL_gain}
\end{center}
\end{figure}

A typical vertical geometry, the so-called vertical cavity surface emitting lasers
(VCSELs),\cite{Chuang:2009,Coldren:2012,Fu:2003,Michalzik:2013} used in nearly 
all spin lasers, is illustrated in Fig.~\ref{fig:VCSEL_gain}(a). Even among 
conventional lasers, VCSELs are recognized for their unique properties, making 
them particularly suitable for optical data transmission.\cite{Michalzik:2013} 
The resonant cavity is usually in the range of the emission wavelength, providing 
a longitudinal single-mode operation. It is formed by a pair of parallel highly 
reflective mirrors made of distributed Bragg reflectors (DBRs), a layered 
structure with varying refractive index. The gain active (gain) region, usually 
consists of III-V quantum wells (QWs) or quantum dots 
(QDs).\cite{Basu2008:APL,Basu2009:PRL,Saha2010:PRB,Oszwaldowski2010:PRB,Lee2012:PRB,Adams2012:IEEEPJ,%
Khaetskii2013:PRL} %I

The key effect of the active region is producing a stimulated emission and coherent 
light that makes the laser such a unique light source. The corresponding optical 
gain that describes stimulated emission, under sufficiently strong pumping/injection 
of carriers, can be illustrated pictorially in Figs.~\ref{fig:VCSEL_gain}(b) 
and ~\ref{fig:VCSEL_gain}(c) for both conventional and spin lasers, respectively. 
In the latter case, it is convenient to decompose the photon density into different 
circular polarizations and distinguish that the gain is generally polarization-dependent. 
If we neglect any losses in the resonant cavity, such a gain would provide an 
exponential growth rate with the distance across a small segment of gain 
material.\cite{Coldren:2012} Since both static and dynamic operations of spin lasers 
depend crucially on their corresponding optical gain, our focus will be to provide 
its microscopic description derived from an accurate electronic structure of an 
active region. 

After this Introduction, in Sec.~II we provide a theoretical framework to calculate 
the gain in quantum well-based lasers. In Sec.~III, we describe the corresponding 
electronic structure and the carrier populations under spin injection, the key prerequisites 
to understanding the spin-dependent gain and its spectral dependence, discussed in Sec.~IV. 
Our gain calculations in Sec.~V explain how the steady-state properties of spin lasers 
can be modified by spin-polarized carriers, carrier density, and resonant cavity 
design. In Sec.~VI, we analyze the influence of a uniaxial strain in the active 
region, which introduces a large birefringence with the resulting oscillation 
frequency that would significantly exceed the frequency range possible in conventional 
lasers. In Sec.~VII, we describe various considerations for the optimized design of 
spin lasers and the prospect of their ultrahigh-frequency operation. A brief 
summary in Sec.~VIII ends our paper.

%===============================================================================

\section{Theoretical framework}

While both QWs and QDs,\cite{Basu2008:APL,Basu2009:PRL, Saha2010:PRB} are used for the active 
region of spin lasers, we focus  here on the QW implementation also  found 
in most of the commercial VCSELs.\cite{Michalzik:2013} To obtain an accurate electronic 
structure in the active region, needed to calculate optical gain, we use the $8 \times 8$ 
$\bm{k{\cdot}p}$ $\bm{k{\cdot}p}$ 
method.\cite{Holub2011:PRB} The total Hamiltonian of the QW system, with 
the growth axis along the $z$ direction, is
\begin{equation}
H_{\textrm{QW}}(z) = H_{kp}(z) + H_{\textrm{st}}(z) + H_{\textrm{O}}(z),
\label{eq:HQW}
\end{equation} 
where $H_{kp}(z)$ denotes the $\bm{k{\cdot}p}$ 
 term, $H_{\textrm{st}}(z)$ describes the 
strain term, and $H_{\textrm{O}}(z)$ includes the band-offset at the interface 
that generates the QW energy profile. The explicit form of these different terms 
for zinc-blende crystals is given in Appendix A.

Because common nonmagnetic semiconductors are well characterized by the 
vacuum permeability, $\mu_0$, a complex dielectric function $\varepsilon(\omega)=\varepsilon_r(\omega)+\varepsilon_i(\omega)$, 
where $\omega$ is the photon (angular) frequency, can be used to simply express 
the dispersion and absorption of electromagnetic waves. The absorption coefficient 
describing gain or loss of the amplitude of an electromagnetic wave propagating 
in such a medium is the negative value of the gain coefficient (or gain 
spectrum),\cite{Haug:2004,Chuang:2009,note_gain} 
\begin{equation}
g^a(\omega)=-\frac{\omega}{c n_r}
\epsilon^a_i(\omega) \, ,
\label{eq:gain}
\end{equation}
where $c$ is the speed of light, $n_r$ is the dominant real part of the refractive 
index of the material,\cite{Haug:2004} and $\varepsilon^a_i(\omega)$ is the imaginary 
part of the dielectric function which generally depends on the polarization of 
light, $a$, given by
\begin{equation}
\varepsilon^a_i(\omega)=C_0\underset{c,v,\vec{k}}{\sum} \left| p^a_{cv\vec{k}} \right|^2 \left(f_{v\vec{k}}-f_{c\vec{k}}\right) 
\delta\left[\hbar\omega_{{cv}\vec{k}}-\hbar\omega\right], 
\label{eq:epsI}
\end{equation}
where the indices $c$ (not to be confused with the speed of light) and $v$ label 
the conduction and valence subbands, respectively, $\vec{k}$ is the wave vector, 
$p^a_{cv\vec{k}}$ is the interband dipole transition amplitude, $f_{c(v)\vec{k}}$ 
is the Fermi-Dirac distribution for the electron occupancy in the conduction (valence) 
subbands, $\hbar$ is the Planck's constant, $\omega_{{cv}\vec{k}}$ is the interband 
transition frequency, and $\delta$ is the Dirac delta-function, which is often 
replaced to include broadening effects for finite lifetimes.\cite{Chuang:2009,Chow:1999} 
The constant $C_0$ is $C_0 = 4\pi^2 e^2/(\varepsilon_0 m_0^2\omega^2\Omega)$,
where $e$ is the electron charge, $m_0$ is the free electron mass, and $\Omega$ 
is the QW volume.

Analogously to expressing the total photon density in Fig.~\ref{fig:VCSEL_gain},
as the sum of different circular polarizations, $S=S^+ +S^-$, in spin-resolved 
quantities we use subscripts to describe different spin projections, i. e., eigenvalues 
of the $\sigma_z$ Pauli matrix. The total electron/hole density can be written as the 
sum of the spin up ($+$) and the spin down ($-$) electron/hole densities, 
$n=n_+ + n_-$ and $p=p_+ + p_-$. In this convention,\cite{Gothgen2008:APL, Lee2010:APL, 
Lee2014:APL} using the conservation of angular momentum between carriers and photons, 
the recombination terms are $n_+p_+$, $n_-p_-$, while the corresponding polarization 
of the emitted light depends on the character of the valence band holes.\cite{note_rate_holes} 
We can simply define the carrier spin polarization
\begin{equation}
P_\alpha = (\alpha_+ - \alpha_-)/(\alpha_+ + \alpha_-),
\label{eq:polar}
\end{equation}
where $\alpha=n$, $p$.\cite{note_polarizations}

Using the dipole selection rules for the spin-conserving interband transitions, 
the gain spectrum, 
\begin{equation}
g^a(\omega) = g^a_+(\omega) + g^a_-(\omega)
\end{equation}
can be expressed in terms of the contributions of spin up and down carriers. To 
obtain $g^a_{+(-)}(\omega)$, the summation of conduction and valence subbands is 
restricted to only one spin: $\underset{c}{\sum} \rightarrow \underset{c+(-)}{\sum}$ 
and $\underset{v}{\sum} \rightarrow \underset{v+(-)}{\sum}$ in Eq.~(\ref{eq:epsI}).

\begin{figure}[h]
\begin{center}
\includegraphics{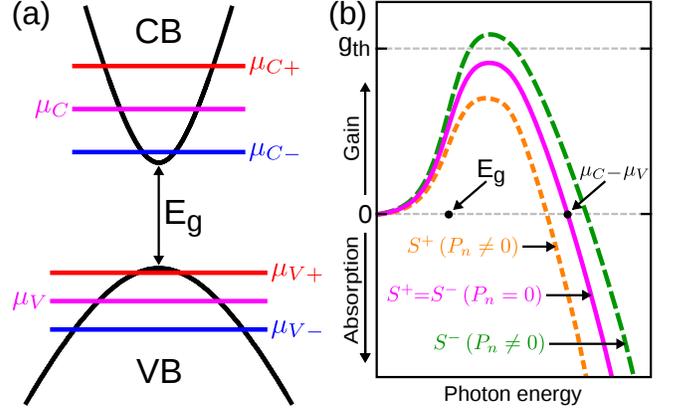}
\caption{(Color online) (a) Energy band diagram with a bandgap $E_g$ and chemical potentials in 
conduction (valence) bands, $\mu_C$ ($\mu_V$) that in the presence of spin-polarized 
carriers become spin-dependent: $\mu_{C(V)+} \neq \mu_{C(V)-}$. Unlike the rest
of our analysis, here holes are spin-polarized. 
(b) Gain spectrum 
for unpolarized (solid) and spin-polarized electrons (dashed curves). 
Positive gain corresponds to the 
emission and negative gain to the absorption of photons. The gain threshold $g_{th}$, 
required for lasing operation, is attained only for $S^-$ helicity of light.} 
\label{fig:chemical_gain}
\end{center}
\end{figure}

To see how spin-polarized carriers could influence the gain, we show chemical potentials, 
$\mu_{C(V)}$, for a simplified conduction (valence) band in Fig.~\ref{fig:chemical_gain}(a). 
The spin imbalance in the active region implies that $\mu_{C(V)}$ 
 will also become spin-dependent. 
Such different chemical potentials lead to the dependence of gain on the polarization 
of light, described in Fig.~\ref{fig:chemical_gain}(b). Without spin-polarized 
carriers, the gain is the same for $S^+$ and $S^-$ helicity of light. In an ideal 
semiconductor laser, $g>0$ requires a population inversion for photon energies, 
$E_g< \hbar \omega < (\mu_C-\mu_V)$. However, a gain broadening is inherent to 
lasers and, as depicted in Fig.~\ref{fig:chemical_gain}(b), $g>0$ even below the 
bandgap, $\hbar \omega<E_g$. If we assume $P_n\neq0$ [recall Eq.~(\ref{eq:polar})] 
and $P_p=0$ (accurately satisfied, as spins of holes relax much faster than electrons), 
we see different gain curves for $S^+$ and $S^-$. The crossover from emission to 
absorption is now in the range of ($\mu_{C-} - \mu_{V-}$) and ($\mu_{C+} - \mu_{V+}$). 

Optical injection of spin-polarized electrons is the most frequently used method 
to introduce spin-imbalance in lasers. Some spin lasers are therefore readily 
available since they can be based on commercial semiconductor lasers to which a 
source of circularly polarized light is added subsequently.\cite{Rudolph2003:APL} 
Such spin injection can be readily understood from dipole optical selection rules which apply for both 
excitation and radiative recombination.\cite{Meier:1984, Zutic2004:RMP}

\begin{figure}[h]
\begin{center}
\includegraphics{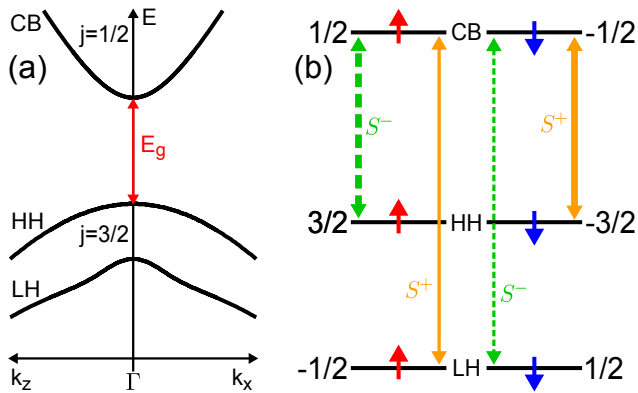}
\caption{(Color online) Schematic band structure and optical selection rules in 
zinc-blende QWs. (a) Conduction band (CB) and valence band with heavy and light 
holes (HH, LH) labeled by their total angular momentum $j=1/2$ and $j=3/2$, representing 
the states of the orbital angular momentum $l=0$ and $1$, respectively (Appendix A). 
(b) Interband dipole transitions near the band edge of a QW for light with positive 
and negative helicity, $S^{\pm}$, between the sublevels labeled by $m_j$, the 
projection of the total angular momentum on the $+z$-axis (along the growth direction). 
The small arrows represent the projection of spin 1/2 of the orbital part that 
contributes to the transition, indicating that dipole transitions do not change 
spin (Appendix B).}
\label{fig:ZB_rules}
\end{center}
\end{figure}

A simplified band diagram for a zinc-blende QW semiconductor with the corresponding 
interband transitions is depicted in Fig.~\ref{fig:ZB_rules}. At the Brillouin zone 
center, the valence band degeneracy of heavy and light holes (HH, LH) in the bulk semiconductor 
is lifted for QWs due to quantum confinement along the growth direction. The angular 
momentum of absorbed circularly polarized light is transferred to the semiconductor. 
Electrons' orbital momenta are directly oriented by light and, through spin-orbit 
interaction, their spins become polarized.\cite{Meier:1984} While initially holes are also polarized, 
their spin polarization is quickly lost.\cite{Zutic2004:RMP}
Thus, as in Fig.~\ref{fig:chemical_gain}(b),  we assume throughout this work $P_p=0$, unless stated otherwise.  

The spin polarization of excited electrons depends on the photon energy for $S^+$ 
or $S^-$ light. From Fig.~\ref{fig:ZB_rules}(b) we can infer that if only CB-HH 
are involved, $|P_n| \rightarrow 1$. At a larger $\hbar \omega$, involving also 
CB-LH transitions, $|P_n|$ is reduced. Expressing $S^\pm\propto Y_1^{\pm1}$, 
where $Y_l^m$ is the spherical harmonic, provides a simple connection between dipole 
selection rules and the conservation of angular momentum in optical transitions 
(Appendix B).

%===============================================================================

\section{Electronic structure}

\begin{figure}[h]
\begin{center}
\includegraphics{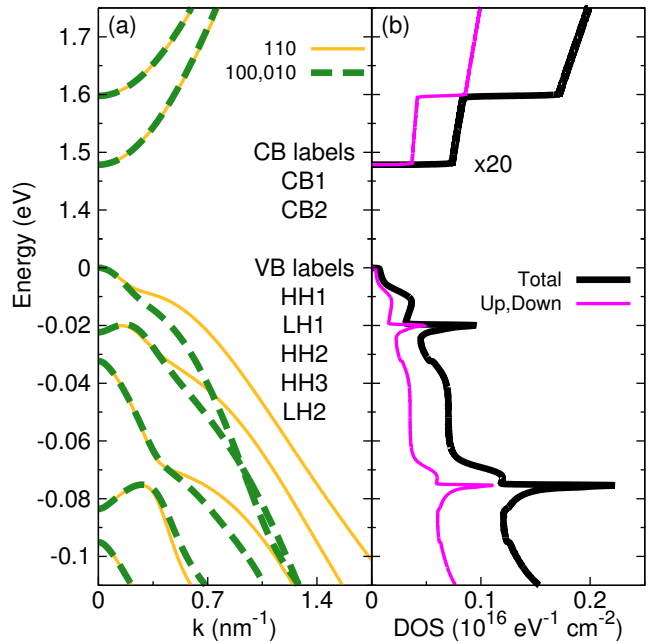}
\caption{(Color online) (a) Band structure for the $\textrm{Al}_{0.3}\textrm{Ga}_{0.7}\textrm{As}$/GaAs 
QW for different $k$-directions along [100],   [010],  and [110]. 
(b) DOS calculated from (a). 
The conduction band DOS is multiplied by a factor of 20 to match the valence band scale. 
The bandgap is $E_g=1.479\; \textrm{eV}$ 
(CB1-HH1 energy difference).}
\label{fig:bs_dos}
\end{center}
\end{figure}

For our microscopic description of spin lasers we focus, on a (Al,Ga)As/GaAs-based 
active region, a choice similar to many commercial VCSELs. We consider an $\textrm{Al}_{0.3}\textrm{Ga}_{0.7}\textrm{As}$ 
barrier and a single 8 nm thick GaAs QW.\cite{note_parameters} The corresponding 
electronic structure of both the band dispersion and the density of states (DOS) is 
shown in Fig.~\ref{fig:bs_dos}. Our calculations, based on the $\bm{k{\cdot}p}$   
method 
and the 8$\times$8 Hamiltonian from Eq.~(\ref{eq:HQW}) (Appendix A), yield two 
confined CB subbands: CB1, CB2, and five VB subbands, labeled in Fig.~\ref{fig:bs_dos}(a) 
by the dominant component of the total envelope function at $\vec{k}=0$. The larger 
number of confined VB subbands stems from larger effective masses for holes than 
electrons.\cite{GaAs_masses} These differences in the effective masses also appear in 
the DOS shown in in Fig.~\ref{fig:bs_dos}(b). 

\begin{figure}[h]
\begin{center}
\includegraphics{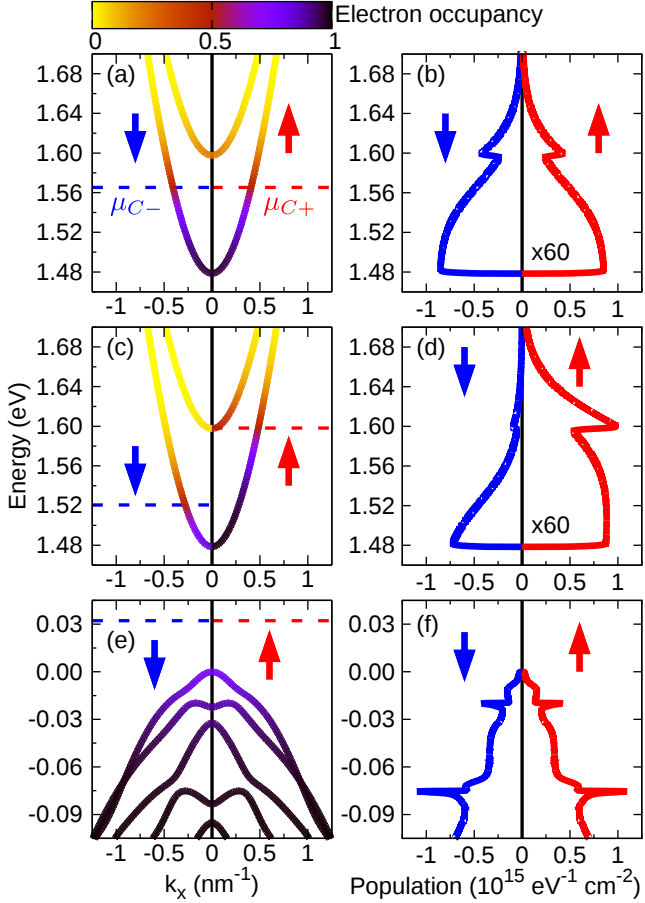}
\caption{(Color online) Band structure of Fig.~\ref{fig:bs_dos}(a) 
with electron occupancy for  (a) $P_n$=0, (c) $P_n=0.5$, and (e) $P_p=0$. Carrier population expressed as a 
product of DOS from  Fig.~\ref{fig:bs_dos}(b) and the Fermi-Dirac distribution of 
electrons for (b) $P_n$=0, (d) $P_n=0.5$, and (f) $P_p=0$. The carrier density is 
fixed at $n=p=3\times10^{12}\;\textrm{cm}^{-2}$ and $T=300\;\textrm{K}$. 
The negative (positive) side of the x-axis represents spin down (up) electrons, dashed 
lines denote chemical potentials. The CB population is multiplied by
60 and shown in the same scale as for the VB.}
\label{fig:bs_occ}
\end{center}
\end{figure}

As we seek to describe the gain spectrum in the active region, once we have the 
electronic structure, it is important to understand the effects associated with 
carrier occupancies though injection/pumping [recall Fig.~2, Eqs.~(2) and (3)]. 
In Figs.~\ref{fig:bs_occ} (a), (c), and (e) we show both examples of injected 
unpolarized ($P_n=0$) and spin-polarized ($P_n=0.5$) electrons as seen in the equal 
and spin-split CB chemical potentials, respectively.
The carrier population\cite{Coldren:2012} is given 
in Figs.~\ref{fig:bs_occ}(b), (d), and (f) using the product of the Fermi-Dirac 
distribution and the DOS for CB and VB for both spin projections.

%===============================================================================

\section{Understanding the spin-dependent gain}

From the conservation of angular momentum and polarization-dependent optical transitions, 
we can understand that even in conventional lasers, 
carrier spin plays a role in determining the gain. 
However, in the absence of spin-polarized carriers\cite{note_nonmagnetic} 
the gain is identical for the two helicities: $g^+=g^-$, and we recover a simple 
description (spin- and polarization-independent) from Fig.~2(b).  In our notation, 
$g^\pm_\pm$,   the upper (lower) index refers to the circular polarization
(carrier spin) [recall Eq.~(5)].

\begin{figure}[h]
\begin{center}
\includegraphics{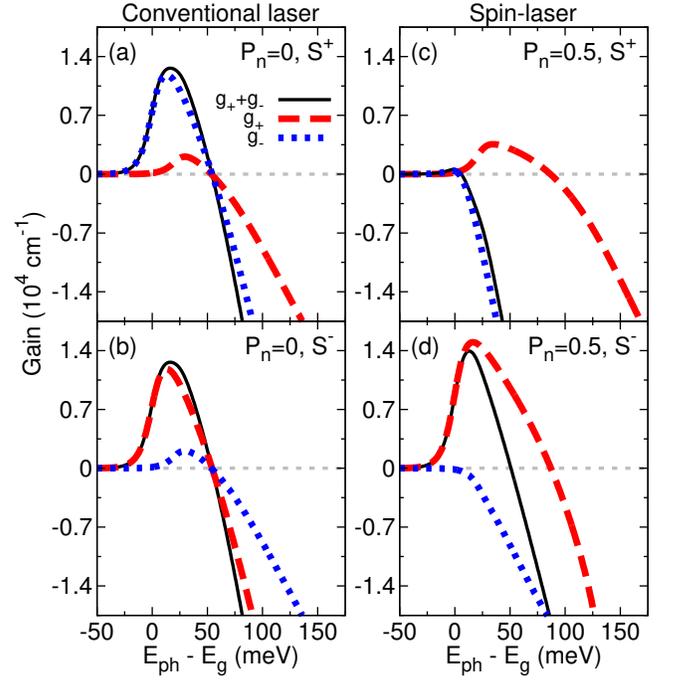}
\caption{(Color online) Gain spectra shown as a function of photon energy measured 
with respect to the energy bandgap. Conventional laser, $P_n=0$ for (a) $S^+$ 
and (b) $S^-$ light polarization. spin lasers, $P_n=0.5$ for (c) $S^+$ and (d) $S^-$ 
light polarizations. The carrier density $n=p=3\times10^{12}\;\textrm{cm}^{-2}$ and 
$T=300\;\textrm{K}$ are the same as in Fig.~\ref{fig:bs_occ}.}
\label{fig:spin_gain}
\end{center}
\end{figure}

This behavior can be further understood from the gain spectrum in Figs.~\ref{fig:spin_gain}(a) 
and (b), where we recognize that $g^+=g^-$ requires: (i) $g^+_- = g^-_+$ and $g^+_+ = g^-_-$, 
dominated by CB1-HH1 ($1.479\;\textrm{eV} = E_g$) 
and CB1-LH1 ($1.501 \; \textrm{eV}$) 
transitions, respectively (recall Fig.~\ref{fig:bs_dos}).  
No spin-imbalance implies 
spin-independent $\mu_C$ and $\mu_V$ [Fig.~3(a)] and thus $g^\pm$, $g^\pm_+$, and 
$g^\pm_-$, all vanish 
the photon energy E$_\textrm{ph}=\hbar \omega = \mu_C - \mu_V$. Throughout our calculations 
we choose a suitable $\cosh^{-1}$ broadening\cite{Chow:1999} with a full width at half-maximum (FWHM) of $19.75\;\textrm{meV}$, 
which accurately recovers the gain of conventional (Al,Ga)As/GaAs QW systems.

We next turn to the gain spectrum of spin lasers. Why is their output different 
for $S^+$ and $S^-$ light, as depicted in Fig.~2(b)? Changing only $P_n=0.5$ from 
Figs.~\ref{fig:spin_gain}(a) and (b), we see very different results for $S^+$ and 
$S^-$ light in Figs.~\ref{fig:spin_gain}(c) and (d). $P_n>0$ implies that 
$\mu_{C+}>\mu_{C-}$ [see Fig.~\ref{fig:bs_occ}(c)], leading to a larger recombination 
between the spin up carriers ($n_+p_+ > n_-p_-$) and thus to a larger $g_+$ for 
$S^+$ and $S^-$ (red/dashed line) than $g_-$ (blue/dashed line). The combined 
effect of having spin-polarized carriers and different polarization-dependent optical 
transitions for spin up and down recombination is then responsible for $g^+ \neq g^-$, 
given by solid lines in Figs.~\ref{fig:spin_gain}(c) and \ref{fig:spin_gain}(d). 
For this case, the emitted light $S^-$ exceeds that with the opposite helicity, $S^+$, 
and there is a gain asymmetry,
%I known also as the gain anisotropy,
\cite{Holub2007:PRL,Hovel2008:APL,Basu2009:PRL} 
another consequence of the polarization-dependent gain. 
The zero gain is attained at $\mu_{C+} - \mu_{V}$ for spin up (red curves) and $\mu_{C-} - \mu_{V}$ 
for spin down transitions (blue curves). The total gain, including both of these 
contributions, reaches zero at an intermediate value. Without any changes to the 
band structure, a simple reversal of the carrier spin-polarization, 
 $P_n\rightarrow -P_n$, reverses the role of  preferential light polarization. 
%===============================================================================

\section{Steady-state gain properties}

Within our framework, providing a spectral information for the gain, we can investigate 
how the carrier density and its spin  polarization, which can be readily modified experimentally, 
can change the steady-state operation of spin lasers. Specific to VCSELs,  
it is important to examine how their laser operation is related to the choice of 
a resonant cavity which defines the photon energy at which the constructive interference 
takes place.

\begin{figure}[h]
\begin{center}
\includegraphics{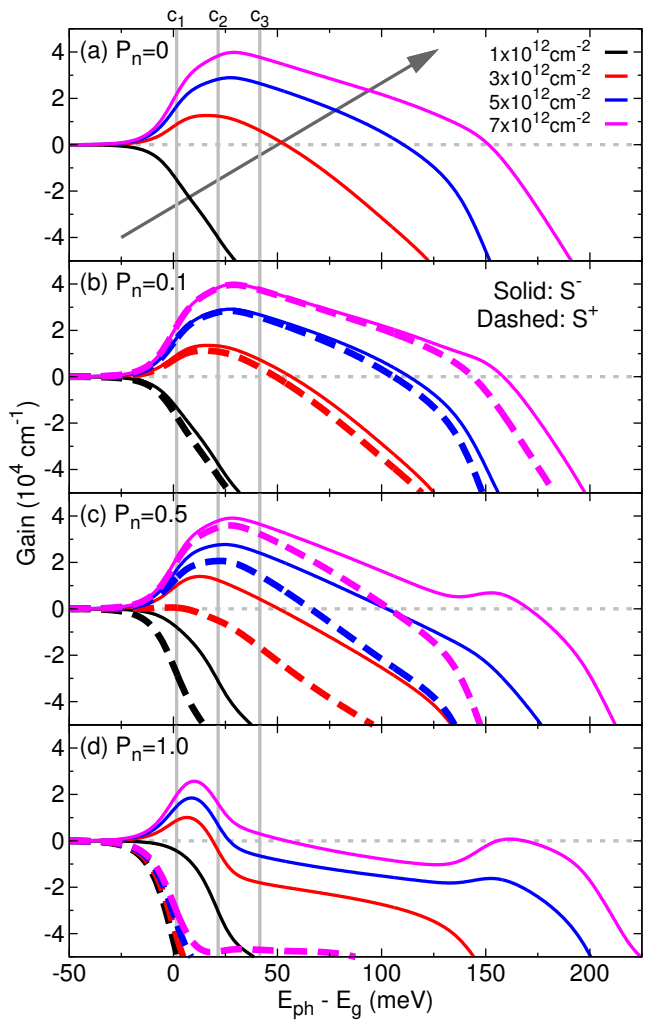}
\caption{(Color online) Evolution of the gain spectra with carrier density for: (a) $P_n=0$, 
(b) $P_n=0.1$, (c) $P_n=0.5$, and (d) $P_n=1.0$. In order to achieve emission, a certain 
value of carrier density should be added to the system. The second peak at
$\textrm{E}_{\textrm{ph}}-\textrm{E}_{\textrm{g}} \sim 150 \; \textrm{meV}$ is 
related to transitions of CB2-HH2. Transitions of CB2-LH2 at $\textrm{E}_{\textrm{ph}}-\textrm{E}_{\textrm{g}} \sim 200 \; \textrm{meV}$ 
can be seen in the broader second peak for $P_n=1.0$. The difference between $g^+$ 
and $g^-$ that arise due to spin-polarized carriers in the system increases with 
$P_n$. For $P_n=1.0$ there is no emission of $S^+$ polarized light, i. e., this 
component is totally absorbed by the system. The diagonal arrow in Fig. 
\ref{fig:gain_spectra} indicates the increase of carrier density in the curves.}
\label{fig:gain_spectra}
\end{center}
\end{figure}

Most of the QW-based lasers do not have a doped active region, and they rely on a charge 
neutral carrier injection (electrical or optical).\cite{Coldren:2012} Here we 
choose $n = p = 1,3,5,7 \times 10^{12} \; \textrm{cm}^{-2}$, and spin polarizations 
$P_n=0, 0.1, 0.5, 1$, respectively. Electrical injection in intrinsic III-V QWs 
using Fe or FeCo allows for $|P_n| \sim 0.3-0.7$,\cite{Hanbicki2002:APL, Zega2006:PRL, 
Salis2005:APL} while $|P_n|\rightarrow 1$ is attainable optically at room 
temperature.\cite{Zutic2004:RMP} In most of the spin lasers, $|P_n| \lesssim 0.2$ 
in the active region. We focus on three resonant cavity positions: $\textrm{c}_1$, $\textrm{c}_2$, 
$\textrm{c}_3$ (vertical lines), defining the corresponding energy of emitted photons
$\textrm{c}_1=1.48 \;\textrm{eV}\sim 1.479$ 
eV (CB1-HH1 transition), $\textrm{c}_2=1.5\;\textrm{eV}\sim 1.501$ eV (CB1-LH1 
transition), and $\textrm{c}_3=1.52\;\textrm{eV}$ (at the high energy side of the 
gain spectrum).

The corresponding results are given in Fig.~\ref{fig:gain_spectra} showing gain 
spectra different for $S^+$ and $S^-$. This gain asymmetry, %I nisotropy, 
$g^+ \neq g^-$,
is more pronounced at larger $P_n$; at $P_n=1$, there is even no $S^+$ emission.
While this trend is expected and could be intuitively understood, there is a more 
complicated dependence of the gain asymmetry,  %I nisotropy, 
$g^-(\hbar\omega)-g^+(\hbar\omega)$
 on the carrier density and the choice 
of the detuning,\cite{Haug:2004} the energy (frequency) difference between the gain 
peak and the resonant cavity position.

\begin{figure}[h]
\begin{center}
\includegraphics{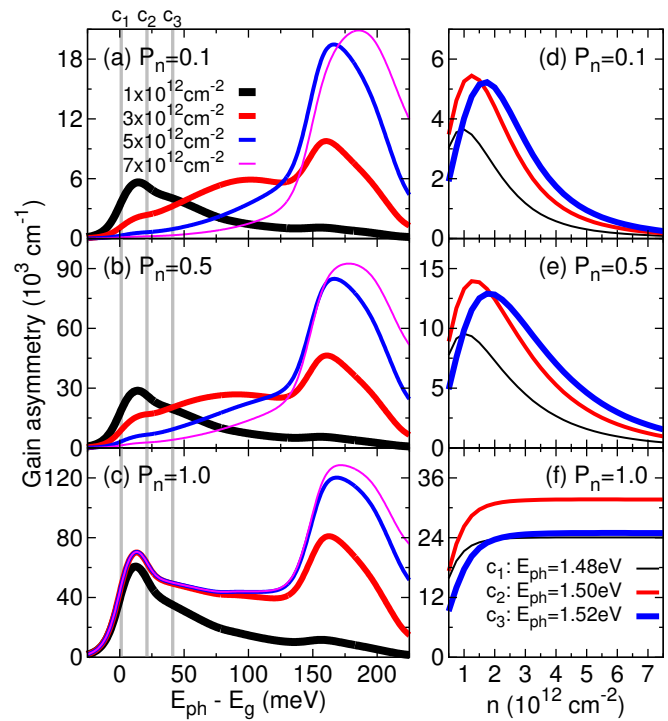}
\caption{(Color online) Gain asymmetry  %I nisotropy 
obtained from Fig.~\ref{fig:gain_spectra} 
for: (a) $P_n=0.1$, (b) $P_n=0.5$, and (c) $P_n=1.0$. As more carriers are added 
to the system, the asymmetry %I nisotropy 
peak shifts to higher energies, however, this energy 
region is not necessarily in the regime of a positive gain. Gain asymmetry %I nisotropy 
as a function of carrier density for: (d) $P_n=0.1$, (e) $P_n=0.5$, and (f) $P_n=1.0$. 
Similar to the case of Figs.~\ref{fig:gain_ani}(a)-(c), the asymmetry %I nisotropy 
peaks may 
not correspond to positive gain.}
\label{fig:gain_ani}
\end{center}
\end{figure}

The gain asymmetry %I nisotropy  
is one of the key figures of merit for spin lasers, and it can be 
viewed as crucial for their spin-selective properties, including robust spin-filtering 
or spin-amplification, in which even a small $P_n$ (few percent) in the active 
region leads to an almost  complete polarization of the emitted light (of just one 
helicity).\cite{Iba2011:APL} Unfortunately, how to enhance the gain asymmetry, %I nisotropy, 
beyond just increasing $P_n$, is largely unexplored.

\begin{figure}[h]
\begin{center}
\includegraphics{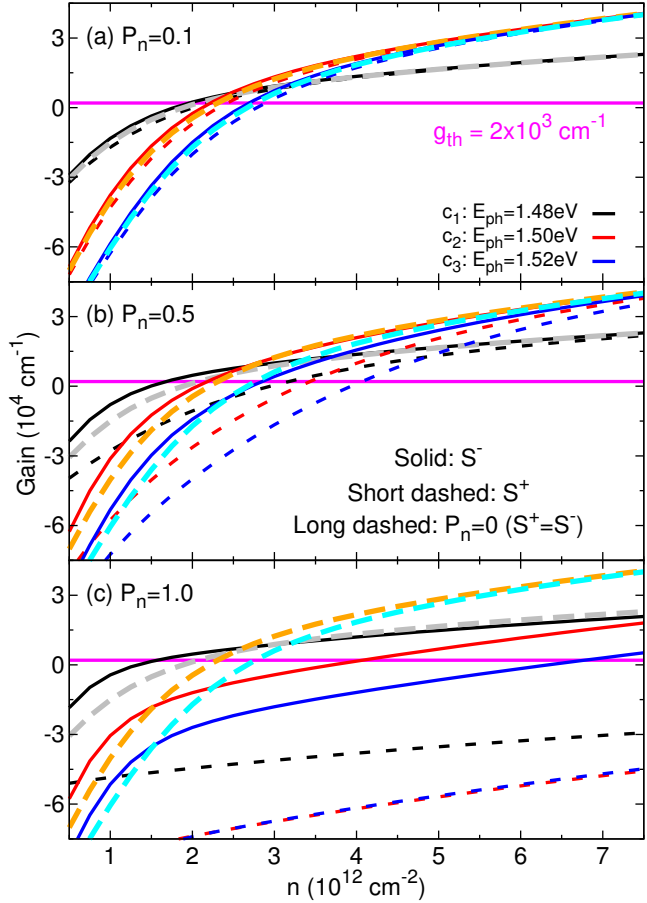}
\caption{(Color online) Gain as a function of carrier density for: (a) $P_n=0.1$, 
(b) $P_n=0.5$, and (c) $P_n=1.0$, with the cavity choices $\textrm{c}_1$, 
$\textrm{c}_2$, and $\textrm{c}_3$. Comparing: (i) solid and short-dashed lines we 
can examine the spin-filtering effect, and (ii) solid and long-dashed curves, we can 
examine the threshold reduction. The solid horizontal line indicates the gain 
threshold, i.e., the losses in the cavity. To achieve the lasing, the value of 
gain must be greater than the gain threshold.}
\label{fig:gain_values}
\end{center}
\end{figure}

To establish a more systematic understanding of a gain asymmetry, %I nisotropy,  
we closely examine
$g^-(\hbar\omega)-g^+(\hbar\omega)$ in Figs.~\ref{fig:gain_ani}(a)-\ref{fig:gain_ani}(c) for different 
$P_n$, carrier densities, and resonant cavities. Increasing $n$, the anisotropy peak 
shifts to higher $\hbar\omega$, indicating an occupation of higher energy subbands. 
However, the absolute anisotropy peak is not always in the emission region. For 
a desirable operation of a spin laser we should seek a large gain anisotropy with 
a positive (and a preferably large) gain. 
Complementary information is given by
Figs.~\ref{fig:gain_ani}(d)-\ref{fig:gain_ani}(f) with a density evolution of $g^- -g^+$ for different 
cavity positions and $P_n$. Again, we see that the gain asymmetry %I nisotropy 
peak can be 
attained outside of the lasing region.

The results in Fig.~\ref{fig:gain_ani} have shown a complex evolution of the gain asymmetry %I nisotropy 
with the cavity position and carrier density. We now repeat a similar 
analysis for the gain itself in Fig.~\ref{fig:gain_values}. The gain calculated 
for two helicities and unpolarized light ($S^+=S^-$), provides a useful guidance 
for the threshold reduction and the spin-filtering effect, invoked in a simple 
bucket model from Fig.~\ref{fig:bucket}.

We first consider $P_n=0.1$ which shows a behavior with an increase in $n$ or, 
equivalently, an increase in injection, that could be expected from the bucket 
model. The threshold value of the gain (the onset of an overflowing bucket), 
$g_\textrm{th}$, is first reached for $S^-$, then for unpolarized light, a sign 
of  threshold reduction, and finally for $S^+$ (a subdominant helicity from the 
conservation of angular momentum and $P_n>0$). Therefore there is a spin-filtering 
interval of $n$ (small, since $P_n$ itself is small) where we expect lasing with 
only one helicity. A similar behavior appears for all the cavity choices $\textrm{c}_1$, 
$\textrm{c}_2$, and $\textrm{c}_3$.

We next turn to $P_n = 0.5$ where $\textrm{c}_1$ shows trends expected both from 
the bucket model and early work on spin lasers.\cite{Rudolph2003:APL, Holub2007:PRL} 
An increase from $P_n=0.1$ to $0.5$ enhances the threshold reduction and the 
spin-filtering interval. However, different cavity positions $\textrm{c}_2$ and $\textrm{c}_3$ reveal a 
different behavior. There is a region where unpolarized light $S^+=S^-$ (long 
dashed lines) yields a greater gain than for $S^-$ (solid lines). For $\textrm{c}_3$ 
the threshold is attained at smaller $n$ for unpolarized light than for negative 
helicity, i.e., there is no threshold reduction.\cite{note_Holub} With $P_n = 1.0$, the threshold 
reduction is only possible for $\textrm{c}_1$. 

These results reinforce the possibility for a versatile spin-VCSEL design by 
a careful choice of the resonant cavity, but they also caution that, depending on the 
given resonant cavity, the usual intuition about the influence of carrier density 
and spin polarization on the laser operation may not be appropriate. 

%===============================================================================

\section{Strain-induced birefringence}
\label{sec:strain}

An important implication of an anisotropic dielectric function is the phenomenon
of birefringence in which the refractive index, and thus the phase velocity of light, 
depends on the polarization of light.\cite{Coldren:2012} 
Due to phase anisotropies in the laser cavity,\cite{note_phase} the emitted frequencies of linearly 
polarized light in the x- and y-directions ($S^x$ and $S^y$) are usually different. 
Such birefringence is often undesired for the operation of 
conventional lasers since it is the origin for the typical complex polarization 
dynamics and chaotic polarization switching behavior in 
VCSELs.\cite{vanExter1997:PRB,SanMiguel1995:PRA,Sondermann2004:IEEE,Al-Seyab2011:IEEEPJ,%
Virte2013:NP} 
While strong values of birefringence are usually considered to be an obstacle for 
the polarization control in spin-polarized lasers,\cite{Hovel2008:APL, Frougier2015:OE} 
the combination of a spin-induced gain asymmetry %I nisotropy 
with birefringence in spin-VCSELs 
allows us to generate fast and controllable oscillations between $S^+$ and $S^-$ 
polarizations.\cite{Gerhardt2011:APL, Hopfner2014:APL} The frequency of these 
polarization oscillations is determined by the linear birefringence in the VCSEL 
cavity, and it can be much higher than the frequency of relaxation oscillations of 
the carrier-photon system in conventional VCSELs. This may pave the way towards 
ultrahigh bandwidth operation for optical communications.\cite{Gerhardt2011:APL, 
Gerhardt2012:AOT, Lee2014:APL}

\begin{figure}[h]
\begin{center}
\includegraphics{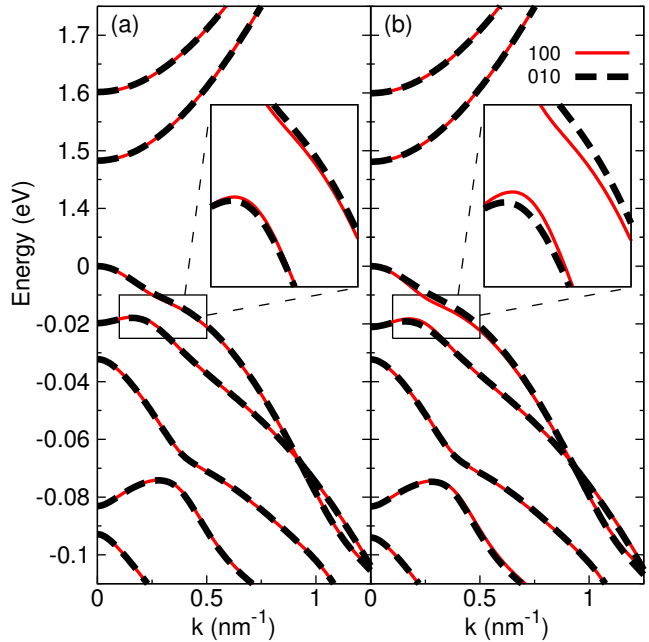}
\caption{(Color online) Band structure with uniaxial strain in the active region 
for (a) $\varepsilon_{xx} \sim 0.019 \%$ and (b) $\varepsilon_{xx} \sim 0.058 \%$. 
The inset shows a zoom around the HH1 and LH1 interaction region, where the difference 
between [100] and [010] directions is more visible. The energy gap of the system 
is $E_g \sim 1.483 \; \textrm{eV}$ for case (a) and $E_g \sim 1.481 \; \textrm{eV}$ 
for case (b).}
\label{fig:bir_bs}
\end{center}
\end{figure}

In order to investigate birefringence effects in the active region of a conventional 
laser, we consider uniaxial strain by extending 
the lattice constant  in x-direction. 
For simplicity, we assume the barrier to have the same lattice constant as GaAs, $5.6533 \; \textrm{\AA}$, 
in y-direction. Therefore, both barrier and well regions will have the same extension 
in x-direction. For $a_x = 5.6544 \; \textrm{\AA}$ we have the corresponding 
element of the strain tensor $\varepsilon_{xx} \sim 0.019 \%$, while $a_x = 5.6566 \; \textrm{\AA}$ 
gives $\varepsilon_{xx} \sim 0.058 \%$. 

The effect of uniaxial strain in the band structure is presented in Fig.~\ref{fig:bir_bs}(a) 
and (b) for $\varepsilon_{xx} \sim 0.019 \%$ and $\varepsilon_{xx} \sim 0.058 \%$, 
respectively. The labeling and ordering of subbands follows the same as that in  
Fig.~\ref{fig:bs_dos}(a). Just this slight anisotropy in the x- and y-lattice constants 
creates a difference in subbands for the [100] and [010] directions. In the inset we 
show the region around the anti-crossing of HH1 and LH1 subbands, where the difference 
is more visible.

\begin{figure}[h]
\begin{center}
\includegraphics{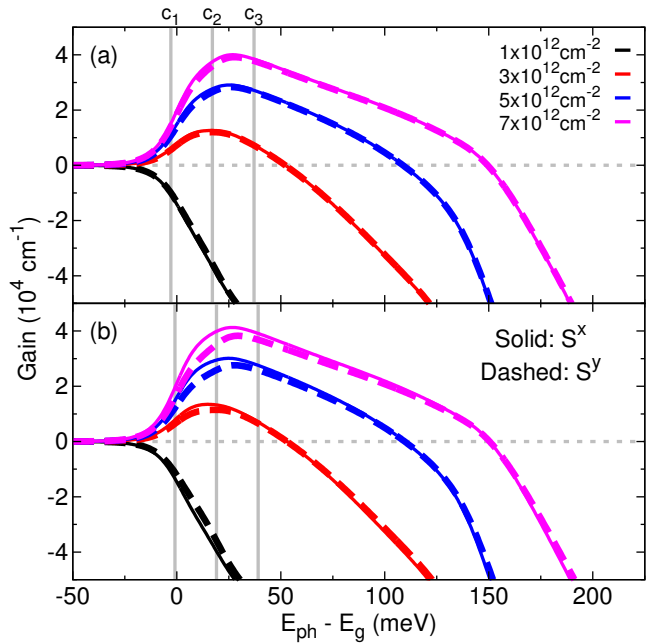}
\caption{(Color online) Uniaxial strain modification of gain spectra for strain 
(a) $\varepsilon_{xx} \sim 0.019 \%$ and (b) $\varepsilon_{xx} \sim 0.058 \%$. The 
anisotropy in the lattice constants for x- and y-directions modifies the output light 
polarization of the laser. Since there are no spin-polarized carriers in the system, 
$g^+ = g^-$.}
\label{fig:bir_gain_spectra}
\end{center}
\end{figure}

In addition to the differences induced in the band structure, the uniaxial strain also 
induces a change in the dipole selection rules between $S^x$ and $S^y$ light polarizations, 
which can be seen in the gain spectra we present in Fig.~\ref{fig:bir_gain_spectra}(a) 
and \ref{fig:bir_gain_spectra}(b) for $\varepsilon_{xx} \sim 0.019 \%$ and $\varepsilon_{xx} \sim 0.058 \%$, 
respectively. Reflecting the features of the band structure, we notice for the 
emission region of the gain spectra that the largest difference between $g^x$ 
and $g^y$ is around the HH1 and LH1 energy regions (between $\textrm{c}_1$ and $\textrm{c}_3$ cavity positions, 
approximately). In the absorption regime (negative gain) we notice $g^x < g^y$,  
while in the emission regime (positive gain) we have $g^x > g^y$. This feature 
is more visible in Fig.~\ref{fig:bir_gain_spectra}(b). 

To calculate the birefringence coefficient in the active region, we used the definition 
of Ref.~\onlinecite{Mulet2002:IEEEJQE}, given by
\begin{equation}
\gamma_{p}(\omega)=-\frac{\omega}{2n_{e}n_{g}}\delta\varepsilon_{r}(\omega) \; ,
\label{eq:gammap}
\end{equation}
where $\omega$ is the frequency of the longitudinal mode in the cavity, 
$n_e$ is the effective index of refraction of the cavity, and $n_g$ is the group refractive 
index. For simplicity, we assume $n_e=n_g$. The real part of the dielectric function 
can be obtained from the imaginary part using the Kramers-Kronig relations.\cite{Haug:2004}

\begin{figure}[h]
\begin{center}
\includegraphics{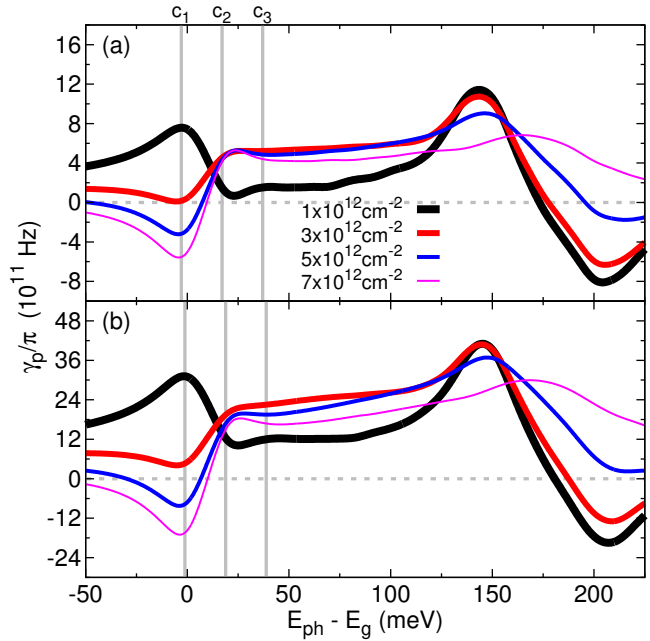}
\caption{(Color online) Birefringence coefficient as a function of photon energy 
considering (a) $\varepsilon_{xx} \sim 0.019 \%$ and (b) $\varepsilon_{xx} \sim 0.058 \%$. 
Just an increase of $0.0022 \; \textrm{\AA}$ in $a_x$ increases $\gamma_p$ by approximately 
3 times. The two peaks,  around $\textrm{E}_{\textrm{ph}}-\textrm{E}_{\textrm{g}} \sim 0$ 
meV and $\textrm{E}_{\textrm{ph}}-\textrm{E}_{\textrm{g}} \sim 150$ meV are related 
to transitions from CB1 and CB2. Transitions related to CB2 are in the absorption 
regime, not visible in Fig.~\ref{fig:bir_gain_spectra}.}
\label{fig:bir_gammap}
\end{center}
\end{figure}

We present the birefringence coefficient in Fig.~\ref{fig:bir_gammap}(a) and \ref{fig:bir_gammap}(b) 
for $\varepsilon_{xx} \sim 0.019 \%$ and $\varepsilon_{xx} \sim 0.058 \%$, respectively. 
We notice that this strain in the active region, responsible for modest
changes in the gain spectra, produces birefringence values of the order of $10^{11-12}$ Hz 
which may be exploited to generate fast polarization oscillations. 
Furthermore, when increasing the strain amount by $\sim 0.04 \%$ from case (a) to case (b), 
the value of $\gamma_p$ increases approximately threefold.\cite{JansenvanDoorn1998:IEEEJQE} 
We also included in 
our calculations spin-polarized electrons and notice that they have only 
a small influence in the birefringence coefficient. Although they change 
$|g^x|$ and $|g^y|$ slightly , the asymmetry %I nisotropy 
is not affected at all for small spin 
polarizations of 10-20\%, which are relevant values in real devices.

\begin{figure}[h]
\begin{center}
\includegraphics{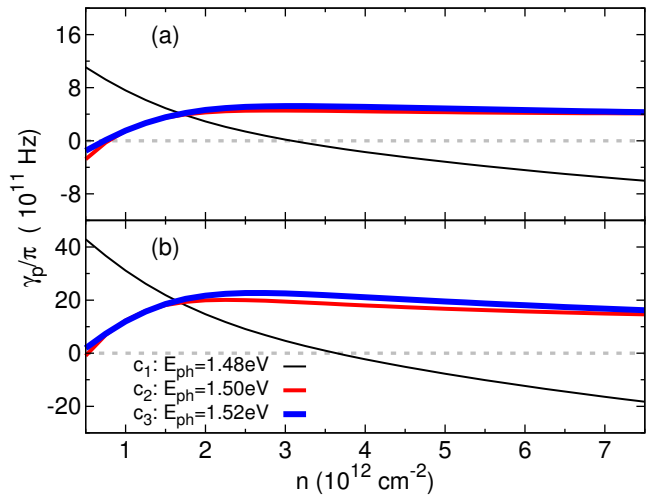}
\caption{(Color online) Birefringence coefficient as function of the carrier density for (a) 
$\varepsilon_{xx} \sim 0.019 \%$ and (b) $\varepsilon_{xx} \sim 0.058 \%$. 
For different cavity designs the behavior of $\gamma_p$ can be completely different.
The carrier density values where $\gamma_p$ changes sign in cavity $\textrm{c}_1$ 
and the flat region in cavities $\textrm{c}_2$ and $\textrm{c}_3$ are already in 
the lasing regime.}
\label{fig:bir_gammap_values}
\end{center}
\end{figure}

Investigating the effect of different cavity designs, we present the values of 
$\gamma_p$ in Figs. \ref{fig:bir_gammap_values}(a) and \ref{fig:bir_gammap_values}(b) 
for $\varepsilon_{xx} \sim 0.019 \%$ and $\varepsilon_{xx} \sim 0.058 \%$, respectively. 
We chose the same photon energies as for the case without birefringence assuming 
that the different values for the strain-induced birefringence in the active region 
will not significantly affect the cavity resonance for reasons of simplicity. For 
the two different strain types the behavior of $\gamma_p$ is very similar for the 
same resonance energy.  
Comparing different cavity designs we observe that for $\textrm{c}_1$, 
the value of $\gamma_p$ strongly decreases and also changes sign with the carrier density, $n$. 
In contrast, for $\textrm{c}_2$ and $\textrm{c}_3$, $\gamma_p$ is always positive. 
After a slow increase with $n$, $\gamma_p$ becomes flat, and nearly independent 
of the carrier density. 

For consistency, we have also calculated the DBR contributions using the approach 
given by Mulet and Balle.\cite{Mulet2002:IEEEJQE} For large anisotropies in the DBR, 
the birefringence coefficient is on the order of $10^{10}$ Hz, consistent with the 
measurements given by van Exter et al.\cite{vanExter1997:PRB} Therefore, for the 
investigated strain conditions, the main contribution to $\gamma_p$ comes from the 
active region and it is a very versatile parameter that can be fine-tuned using 
both carrier density and cavity designs, possibly even changing its sign and reaching 
carrier density-independent regions.

%===============================================================================

\section{Ultra high-frequency operation}
\label{sec:ultra}

Lasers could provide the next generation of parallel optical interconnects and 
optical information processing.\cite{Fu:2003, Coldren:2012, Michalzik:2013, Kaminow:2013, 
Agrawal:2002,Ciftcioglu2012:OE,Miller2009:PIEEE} The growth in communication\cite{Hilbert2011:S} 
and massive data centers\cite{report:2012} will pose further limitations on 
interconnects.\cite{itrs:2011} Conventional metallic interconnects used in multicore 
microprocessors are increasingly recognized as the bottleneck in maintaining Moore's 
law scaling and the main source of power dissipation.\cite{Miller2009:PIEEE,itrs:2011} 
Optical interconnects can effectively address the related limitations, such as 
the electromagnetic crosstalk and signal distortion, while providing a much larger 
bandwidth.\cite{Ciftcioglu2012:OE,Miller2009:PIEEE} VCSELs are considered particularly 
suitable for short-haul communication and on-chip interconnects.\cite{Michalzik:2013} 
However, to fully utilize their potential, it would be important to explore the 
paths for their high-frequency operation and achieve a higher modulation bandwidth, 
limited for conventional lasers to about $\sim 50$ GHz.\cite{Michalzik:2013, Westbergh2013:EL}

How can we understand the frequency limitation of a laser? Why would a higher 
frequency modulation lead to a decrease in a signal-to-noise ratio and limit the 
effective bandwidth? An accurate analogy is provided by a driven and damped harmonic 
oscillator. The laser response, just like the harmonic oscillator, is unable to 
follow a high enough modulation frequency. A Lorenzian-like frequency-dependent 
displacement of a harmonic oscillator closely matches a modulation response of a 
laser, decreasing as $1/\omega$, above the corresponding resonance frequency, known 
as the relaxation oscillation frequency, $\omega_R$, representing a natural oscillation 
between the carriers and photons and often used to estimate the bandwidth of a 
laser.\cite{Coldren:2012, Michalzik:2013, Kapon:1999} 

To realize a high-speed operation in conventional lasers requires a careful design 
and optimization of many parameters. Attaining a high $\omega_R$ is closely related 
to optimizing the gain which increases with $n$,\cite{note_dope} but decreases 
with photon density $S$, 
known as the gain compression\cite{note_compress} which 
would be desirable to minimize. For a small-signal modulation $S(t)=S_0+\delta S(t)$, 
above the threshold,\cite{Coldren:2012} 
\begin{equation}
\omega^2_R\approx v_g(dg/dn)S_0/\tau_\textrm{ph},
\label{eq:omegaR}
\end{equation}
where $v_g$ is the group velocity of the relevant mode, $dg/dn$ is the differential 
gain at the threshold, and $\tau_\textrm{ph}$ is the photon lifetime. While $\omega_R$ 
increases with $S_0$, a larger $S_0$, through gain compression, is detrimental 
by diminishing the differential gain. There are additional factors, beyond Eq.~(\ref{eq:omegaR}), 
required for a high $\omega_R$, such as minimizing the transport time for carriers 
to reach the active region, achieving a high carrier escape rate into the QW barriers, 
and minimizing extrinsic parasitic effects between the intrinsic laser and the 
driving circuit.\cite{Michalzik:2013,Kapon:1999}

Introducing spin-polarized carriers offers additional possibilities to enhance 
$\omega_R$, corresponding to the modulation of the emitted $S$, beyond the frequencies 
attainable in conventional lasers. In the regime of small-signal modulation, 
both $\omega_R$ and the bandwidth have been shown to increase with an increase 
of the spin-polarization of the injected carriers, $P_J$,\cite{Lee2010:APL, Lee2012:PRB} 
associated with the threshold reduction [thus for a given injection $S_0$ is larger 
than in Eq.~(\ref{eq:omegaR})]. Similar trends are predicted in the large-signal 
modulation, but the corresponding increase of $\omega_R$ (as compared to the 
conventional lasers) can exceed what would be expected based only on the threshold 
reduction due to $P_J \neq 0$.\cite{Lee2014:APL}

Another  approach to achieve a higher $\omega_R$ is to use the polarization 
dynamics, instead of the intensity dynamics of the emitted light. The coupling 
between spin-polarized carriers and the light polarization in birefringent  microcavities 
corresponds to different resonant mechanisms than those that govern the light intensity 
and thus to potentially higher $\omega_R$. Early experiments on polarization 
dynamics in VCSELs of Oestreich and collaborators have demonstrated spin-carrier 
dynamics of 120 GHz.\cite{Oestreich:2001} However, their  (Ga,In)As 
QW spin lasers 
operated at 10 K and required a large magnetic field for  
fast spin precession.

Could we attain similar ultrahigh frequencies at room temperature without an applied 
magnetic field? Our findings from Sec.~\ref{sec:strain} suggest that indeed 
such an operation could be realized by a careful design of birefringent cavity 
properties providing frequency splitting of the two orthogonal linearly-polarized 
lasing modes. While in conventional VCSEL only one linearly-polarized mode is emitted, 
injecting spin-polarized carriers leads to the circularly-polarized emission and 
thus the operation of both linearly-polarized modes at the same time. The beating 
between the two frequency-split linearly-polarized modes creates polarization 
oscillations with frequency determined by the birefringence rate, 
$\gamma_p/\pi$.\cite{Gerhardt2011:APL,Hopfner2014:APL}

Strain-induced values of $\gamma_p$ in the active region shown in Figs.~\ref{fig:bir_gammap}
and \ref{fig:bir_gammap_values} are 
sufficiently high to exceed the highest available frequency operation of 
conventional VCSELs. A strong spectral dependence of $\gamma_p$, including a possible 
sign change, requires a careful analysis of the detuning behavior, but it also provides 
important opportunities for desirable operation of spin lasers. For example, a 
large $\gamma_p$ can be achieved with a very weak dependence on the carrier density. 
The feasibility of a high-birefringence rate is further corroborated by the experiments 
using mechanical strain attaining $\gamma_p/\pi \sim 80$ GHz,\cite{Panajotov2000:APL} 
while theoretical calculations suggest even $\gamma_p/\pi \sim 400$ GHz with asymmetric 
photonic crystals.\cite{Dems2008:OC}

%===============================================================================

\section{Conclusions}
\label{sec:conclude}

Our microscopic model of optical gain is based on a similar framework previously 
employed for conventional lasers\cite{Chuang:2009,Coldren:2012,Chow:1999} to simply 
elucidate how introducing spin-imbalance could enable their improved dynamical 
operation. In contrast to the common understanding that the birefringence is detrimental 
for lasers, we focus on the regime of a large strain-induced birefringence to overcome 
frequency limitations in conventional lasers.

With a goal to maximize the birefringence-dominated bandwidth in a experimentally 
realized spin laser, we can use the guidance from the analysis of both high-speed 
conventional lasers and the steady-state operation of spin lasers to explore potential 
limiting factors. Future calculations should also examine the influence of a 
spin-dependent gain compression, Coulomb interactions,\cite{Chow:1999,Burak2000:PRA,Sanders1996:PRB}
an active region with multiple QWs,\cite{Michalzik:2013}, spin relaxation~\cite{Lee2014:APL,Zutic2004:RMP,Zutic2003:APL} 
and a careful analysis of the optimal cavity position that would combine high 
(differential) gain, high-gain asymmetry,%I  nisotropy, 
and high $\gamma_p$.

While currently the most promising path to demonstrate our predictions for ultrahigh 
frequency operation is provided by optically injected spin-polarized carriers to 
the existing VCSELs, there are encouraging developments for electrically injected 
spin-polarized carriers. A challenge is to overcome a relatively large separation 
between a ferromagnetic spin injector and an active region ($> \mu$m) implying 
that at 300 K recombining carriers would have only a negligible spin polarization.\cite{Soldat2012:APL} 
However, room temperature electrical injection of spin-polarized carriers has already 
been realized through spin-filtering by integrating nanomagnets with the active 
region of a VCSEL.\cite{Cheng2014:NN} Additional efforts focus on vertical external 
cavity surface emitting lasers (VECSELs),\cite{Frougier2015:OE,Frougier2013:APL} 
which could enable depositing a thin-film ferromagnet to be deposited just 100-200 nm away from 
the active region, sufficiently close to attain a considerable spin polarization 
of carriers in the active region at room temperature.

An independent progress in spintronics to store and sense information using magnets 
with a perpendicular anisotropy\cite{note_harddrive} and to attain fast magnetization 
reversal\cite{Garzon2008:PRB} could also be directly beneficial for spin lasers. 
Electrical spin injection usually relies on magnetic thin films with in-plane 
anisotropy requiring a large applied magnetic field to achieve an out-of-plane 
magnetization and the projection of injected spin compatible with the carrier 
recombination of circularly polarized light in a VCSEL geometry (along the z-axis, 
see Fig.~\ref{fig:ZB_rules}). 
However, a perpendicular anisotropy could provide an elegant spin 
injection in remanence,\cite{Sinsarp2007:JJAP,Hovel2008:APLa,Zarpellon2012:PRB}
avoiding the technologically undesirable applied magnetic field. The progress in 
fast magnetization reversal could stimulate implementing all-electrical schemes 
for spin modulation in lasers that were shown to yield an enhanced bandwidth in 
lasers.\cite{Gerhardt2011:APL,Hopfner2014:APL,Lee2010:APL,Lee2012:PRB,Lee2014:APL,%
Banerjee2011:JAP,Nishizawa2014:APL}

{\it Note added in proof}. After this work was completed and
submitted, our predictions for high-frequency birefringence
were experimentally demonstrated in similar GaAs/AlGaAs
quantum well spin VCSELs revealing values of $\sim 250$ GHz\cite{Pusch2015:EL}.

%===============================================================================

\section*{Acknowledgements}

We thank M.~R. Hofmann and R. Michalzik for valuable discussions about the feasibility 
of the proposed spin lasers and the state-of-the art active regions in conventional 
lasers. We thank B. Scharf  for carefully reading this manuscript. 
This work has been supported by  CNPq (grant No. 246549/2012-2), FAPESP (grants 
No. 2011/19333-4, No. 2012/05618-0 and No. 2013/23393-8), NSF ECCS-1508873, NSF ECCS-1102092, 
U.S. ONR N000141310754, NSF DMR-1124601, and the German Research Foundation (DFG) 
grant "Ultrafast Spin Lasers for Modulation Frequencies in the 100 GHz Range" GE 
1231/2-1.

%===============================================================================

\setcounter{equation}{0}
\renewcommand{\theequation}{A-\arabic{equation}} 
\section*{Appendix A}

The versatility of the $\bm{k{\cdot}p}$ method has been successfully used to obtain the 
gain spectra in conventional lasers,\cite{Chuang:2009,Coldren:2012,Fu:2003,Chow:1999,Haug:2004}
as well as to elucidate a wealth of other phenomena, such as the spin Hall effect, 
topological insulators, and Zitterbewegung.\cite{Murakami2003:S,Hasan2010:RMP,Bernardes2007:PRL} 
Our own implementation of the $\bm{k{\cdot}p}$ method in this work has been previously
tested in calculating the luminescence spectra in $\delta$-doped GaAs,\cite{Sipahi1998:PRB}  
confirming experimental and theoretical electronic structure for GaAs QWs,\cite{Lee2014:PRB}
and (Al,Ga)N/GaN superlattices,\cite{Rodrigues2000:APL} identifying fully spin-polarized 
semiconductor heterostructures, based on (Zn,Co)O,\cite{Marin2006:APL} and
exploring polytypic systems consisting of zinc-blende and wurtzite crystal phases 
in the same nanostructure.\cite{FariaJunior2012:JAP, FariaJunior2014:JAP}

Before considering confined systems, it is important to investigate the corresponding 
bulk crystal structure and construct the functional form of the Hamiltonian. For 
zinc-blende crystals, the bulk basis set that describes the lower conduction and 
top valence bands is\cite{Zutic2004:RMP,Sipahi1996:PRB,Enderlein:1997,Winkler:2003,Enderlein1997:PRL} 
\begin{eqnarray}
\left|\text{CB}\Uparrow\right\rangle   & = &  \left|S\uparrow\right\rangle \nonumber \\
\left|\text{CB}\Downarrow\right\rangle & = &  \left|S\downarrow\right\rangle \nonumber \\
\left|\text{HH}\Uparrow\right\rangle   & = &  \left|\left(X+iY\right)\uparrow\right\rangle / \sqrt{2} \nonumber \\
\left|\text{LH}\Uparrow\right\rangle   & = & i\left|\left(X+iY\right)\downarrow-2Z\uparrow\right\rangle / \sqrt{6} \nonumber \\
\left|\text{LH}\Downarrow\right\rangle & = &  \left|\left(X-iY\right)\uparrow+2Z\downarrow\right\rangle / \sqrt{6} \nonumber \\
\left|\text{HH}\Downarrow\right\rangle & = & i\left|\left(X-iY\right)\downarrow\right\rangle / \sqrt{2} \nonumber \\
\left|\text{SO}\Uparrow\right\rangle   & = &  \left|\left(X+iY\right)\downarrow+Z\uparrow\right\rangle / \sqrt{3} \nonumber \\
\left|\text{SO}\Downarrow\right\rangle & = & i\left|-\left(X-iY\right)\uparrow+Z\downarrow\right\rangle / \sqrt{3} \, ,
\label{eq:basiskp}
\end{eqnarray}
where, compared to Fig.~\ref{fig:ZB_rules}(a),  we also introduce the spin-orbit spin-split-off 
subbands $\left|\text{SO}\right\rangle$. Here $\left|S\right\rangle$ and $\left|X\right\rangle,\left|Y\right\rangle,\left|Z\right\rangle$ 
are the basis states for irreducible representations $\Gamma_{1} \sim x^2+y^2+z^2$ 
and $\Gamma_{15} \sim x,y,z$, having an orbital angular momentum $l=0$ and $l=1$,  
respectively. The single arrows ($\uparrow,\downarrow$) represent the projection 
of spin angular momentum $s=1/2$ on the $+z$-axis while the double arrows ($\Uparrow,\Downarrow$) 
represent the projection of total angular momentum on the $+z$-axis. Rewriting 
the basis set (\ref{eq:basiskp}) in terms of the total angular momentum $j$ and 
its projection $m_j$, $\left|j, m_j \right\rangle$, we have
\begin{eqnarray}
\left|\text{CB}\Uparrow\left(\Downarrow\right)\right\rangle & = & \left|1/2, \, 1/2 \, \left(-1/2\right)\right\rangle \nonumber \\
\left|\text{HH}\Uparrow\left(\Downarrow\right)\right\rangle & = & \left|3/2, \, 3/2 \, \left(-3/2\right)\right\rangle \nonumber \\
\left|\text{LH}\Uparrow\left(\Downarrow\right)\right\rangle & = & \left|3/2, \, 1/2 \, \left(-1/2\right)\right\rangle \nonumber \\
\left|\text{SO}\Uparrow\left(\Downarrow\right)\right\rangle & = & \left|1/2, \, 1/2 \, \left(-1/2\right)\right\rangle \, .
\end{eqnarray}

In the basis set of Eq.~(\ref{eq:basiskp}), the $\bm{k{\cdot}p}$ term in Eq.~(\ref{eq:HQW}) is 
\begin{widetext}
\begin{equation}
H_{kp}=\left[\begin{array}{cccccccc}
U & 0 & iP_{+} & \sqrt{\frac{2}{3}}P_{z} & \frac{i}{\sqrt{3}}P_{-} & 0 & \frac{i}{\sqrt{3}}P_{z} & \sqrt{\frac{2}{3}}P_{-}\\
0 & U & 0 & -\frac{1}{\sqrt{3}}P_{+} & i\sqrt{\frac{2}{3}}P_{z} & -P_{-} & i\sqrt{\frac{2}{3}}P_{+} & -\frac{1}{\sqrt{3}}P_{z}\\
-iP_{-} & 0 & Q & S & R & 0 & \frac{i}{\sqrt{2}}S & -i\sqrt{2}R\\
\sqrt{\frac{2}{3}}P_{z} & -\frac{1}{\sqrt{3}}P_{-} & S^{\dagger} & T & 0 & R & -\frac{i}{\sqrt{2}}\left(Q-T\right) & i\sqrt{\frac{3}{2}}S\\
-\frac{i}{\sqrt{3}}P_{+} & -i\sqrt{\frac{2}{3}}P_{z} & R^{\dagger} & 0 & T & -S & -i\sqrt{\frac{3}{2}}S^{\dagger} & -\frac{i}{\sqrt{2}}\left(Q-T\right)\\
0 & -P_{+} & 0 & R^{\dagger} & -S^{\dagger} & Q & -i\sqrt{2}R^{\dagger} & -\frac{i}{\sqrt{2}}S^{\dagger}\\
-\frac{i}{\sqrt{3}}P_{z} & -i\sqrt{\frac{2}{3}}P_{-} & -\frac{i}{\sqrt{2}}S^{\dagger} & \frac{i}{\sqrt{2}}\left(Q-T\right) & i\sqrt{\frac{3}{2}}S & i\sqrt{2}R & \frac{1}{2}\left(Q+T\right)-\Delta_{SO} & 0\\
\sqrt{\frac{2}{3}}P_{+} & -\frac{1}{\sqrt{3}}P_{z} & i\sqrt{2}R^{\dagger} & -i\sqrt{\frac{3}{2}}S^{\dagger} & \frac{i}{\sqrt{2}}\left(Q-T\right) & \frac{i}{\sqrt{2}}S & 0 & \frac{1}{2}\left(Q+T\right)-\Delta_{SO}
\end{array}\right] \, ,
\label{eq:Hkp}
\end{equation}
\end{widetext}
with elements
\begin{eqnarray}
Q & = & -k_{x}\left(\tilde{\gamma}_{1}+\tilde{\gamma}_{2}\right)k_{x}-k_{y}\left(\tilde{\gamma}_{1}+\tilde{\gamma}_{2}\right)k_{y}-k_{z}\left(\tilde{\gamma}_{1}-2\tilde{\gamma}_{2}\right)k_{z}\nonumber \\
T & = & -k_{x}\left(\tilde{\gamma}_{1}-\tilde{\gamma}_{2}\right)k_{x}-k_{y}\left(\tilde{\gamma}_{1}-\tilde{\gamma}_{2}\right)k_{y}-k_{z}\left(\tilde{\gamma}_{1}+2\tilde{\gamma}_{2}\right)k_{z}\nonumber \\
S & = & i\sqrt{3}\left[\left(k_{x}\tilde{\gamma}_{3}k_{z}+k_{z}\tilde{\gamma}_{3}k_{x}\right)-i\left(k_{y}\tilde{\gamma}_{3}k_{z}+k_{z}\tilde{\gamma}_{3}k_{y}\right)\right]\nonumber \\
R & = & -\sqrt{3}\left[\left(k_{x}\tilde{\gamma}_{2}k_{x}-k_{y}\tilde{\gamma}_{2}k_{y}\right)-i\left(k_{x}\tilde{\gamma}_{3}k_{y}+k_{y}\tilde{\gamma}_{3}k_{x}\right)\right]\nonumber \\
U & = & E_{g}+k_{x}Ak_{x}+k_{y}Ak_{y}+k_{z}Ak_{z}\nonumber \\
P_{\pm} & = & (1/2\sqrt{2})\left[P\left(k_{x}\pm ik_{y}\right)+\left(k_{x}\pm ik_{y}\right)P\right]\nonumber \\
P_{z} & = & (1/2)\left(Pk_{z}+k_{z}P\right) \, ,
\label{eq:Hkpterms}
\end{eqnarray}
where $\tilde{\gamma}_1$,
$\tilde{\gamma}_2,\tilde{\gamma}_3$, and $A$, given in units of 
$\hbar^{2}/2m_{0}$, are the effective mass parameters of the valence and conduction 
bands, respectively, explicitly given below. 
 The gap is 
 $E_g$, the spin-orbit splitting at the
$\Gamma$ point is 
$\Delta_{SO}$, and $P$ is the Kane parameter of 
the interband interaction, defined as
\begin{equation}
P  =  -i \frac{\hbar}{m_0} \left\langle \alpha \left| p_{l} \right| S \right\rangle \, ,
\end{equation}
with $\alpha=X,Y,Z$ and $l=x,y,z$.

The formulation of a bulk  $\bm{k{\cdot}p}$ model can vary significantly in its complexity, 
the choice of the specific system, and the number of bands included.
In the description of zinc-blende structures, usually either 6$\times$6 or 8$\times$8 models
are employed.\cite{Winkler:2003} 
In the first case, the information of the valence and conduction 
band is decoupled, while in the second case their coupling is explicitly included.
Their effective mass parameters are connected by
\begin{eqnarray}
\tilde{\gamma}_{1} & = & \gamma_{1}-E_P/3E_g \nonumber \\
\tilde{\gamma}_{2} & = & \gamma_{2}-E_P/6E_g \nonumber \\
\tilde{\gamma}_{3} & = & \gamma_{3}-E_P/6E_g \nonumber \\
A & = & \frac{1}{m_{e}^{*}}-\left(\frac{E_{g}+\frac{2}{3}\Delta_{SO}}{E_{g}+\Delta_{SO}}\right)\frac{E_{P}}{E_{g}} \nonumber \\
E_{P} & = & 2m_0 P^2/\hbar^2 \, ,
\label{eq:Hkp86}
\end{eqnarray}
where $\tilde{\gamma}_{1,2,3}$ are used in the 8$\times$8 model and $\gamma_{1,2,3}$ 
in the 6$\times$6 model, 
which can also be related to the tight-binding parameters.\cite{Lee2014:PRB}
To recover the 6$\times$6 model from the 8$\times$8 model, we set $P=0$ in Eqs.~(\ref{eq:Hkp}), (\ref{eq:Hkpterms}) 
and (\ref{eq:Hkp86}).

The strain term, $H_{\textrm{st}}$, takes a form similar to Eq.~(\ref{eq:Hkp}) but 
without the $E_g$, $\Delta_{SO}$ and $P$ parameters. The matrix elements can be written 
as
\begin{eqnarray}
Q_{\textrm{st}} & = & -a_{v}\left(\varepsilon_{xx}+\varepsilon_{yy}+\varepsilon_{zz}\right)-\frac{b}{2}\left(\varepsilon_{xx}+\varepsilon_{yy}-2\varepsilon_{zz}\right)\nonumber \\
T_{\textrm{st}} & = & -a_{v}\left(\varepsilon_{xx}+\varepsilon_{yy}+\varepsilon_{zz}\right)+\frac{b}{2}\left(\varepsilon_{xx}+\varepsilon_{yy}-2\varepsilon_{zz}\right)\nonumber \\
S_{\textrm{st}} & = & d\left(\varepsilon_{yz}+i\varepsilon_{xz}\right)\nonumber \\
R_{\textrm{st}} & = & -\frac{\sqrt{3}b}{2}\left(\varepsilon_{xx}-\varepsilon_{yy}\right)+id\varepsilon_{xy}\nonumber \\
U_{\textrm{st}} & = & a_{c}\left(\varepsilon_{xx}+\varepsilon_{yy}+\varepsilon_{zz}\right) \, ,
\end{eqnarray}
with $a_{v}$, $b$, and $d$ representing the deformation potentials for the valence 
band and $a_c$ for the conduction band. The strain tensor components are given by
$\varepsilon_{ij}\;(i,j=x,y,z)$.

In order to treat a QW system, which now lacks translational symmetry along the 
growth direction, we can replace the exponential part of the Bloch's theorem by 
a generic function. This procedure is called the envelope function approximation\cite{Winkler:2003} 
and it leads to the dependence along the growth direction of the $\bm{k{\cdot}p}$ and strain 
parameters in Hamiltonian terms $H_{kp}(z)$ and $H_{\textrm{st}}(z)$. Also, the 
band-offset at the interface of different materials is taken into account in the 
term $H_{\textrm{O}}(z)$
\begin{equation}
H_{\textrm{O}}(z) = \textrm{diag}\left[ \delta_V(z), \, \cdots, \, \delta_V(z), \, \delta_C(z), \, \delta_C(z) \right] \, ,
\end{equation}
where $\delta_{V(C)}(z)$ describes the energy change in the valence (conduction) band.

Under the envelope function approximation, the QW Hamiltonian from Eq.~(\ref{eq:HQW})
is now described by a system of 8 coupled differential equations that does not generally
have analytical solutions. We solve these equations 
numerically using the plane-wave expansion 
for the $z$-dependent parameters and envelope functions. Details of the 
envelope function approximation and plane wave expansion for QW systems can be 
found in references~\onlinecite{Sipahi1996:PRB,FariaJunior2012:JAP,FariaJunior2014:JAP}.

%===============================================================================

\setcounter{equation}{0}
\renewcommand{\theequation}{B-\arabic{equation}} 
\section*{Appendix B}

The interband dipole transition amplitude that appears in Eq.~(\ref{eq:epsI}) is 
given by
\begin{equation}
\ensuremath{p_{cv\vec{k}}^{a}}=\left\langle c,\vec{k}\left|\hat{a}\cdot\vec{p}\right|v,\vec{k}\right\rangle \, ,
\end{equation}
and for the light polarization $S^{\pm}$ we have
\begin{equation}
\hat{a}=\frac{1}{\sqrt{2}}\left(\hat{x} \pm i\hat{y}\right) \, ,
\end{equation}
and therefore
\begin{equation}
\hat{a}\cdot\vec{p}=\frac{p_{x}\pm ip_{y}}{\sqrt{2}} \, .
\end{equation}

In the simplified QW of Fig.~\ref{fig:ZB_rules}, we are showing the selection rules
for $\vec{k}=0$ and assuming the conduction band as $\left|c,0\right\rangle =\left|\text{CB}\Uparrow\left(\Downarrow\right)\right\rangle $,
and valence band as, $\left|v,0\right\rangle =\left|\text{HH}\Uparrow\left(\Downarrow\right)\right\rangle $
or $\left|v,0\right\rangle =\left|\text{LH}\Uparrow\left(\Downarrow\right)\right\rangle $. 
Calculating the matrix elements between these states, we obtain 

\begin{widetext}
\begin{eqnarray}
\left\langle \textrm{CB}\Uparrow\left|p_{\pm}\right|\textrm{HH}\Uparrow\right\rangle  & = & \left\langle S\uparrow\left|\frac{p_{x}\pm ip_{y}}{\sqrt{2}}\right|\frac{1}{\sqrt{2}}\left(X+iY\right)\uparrow\right\rangle 
 =  \frac{1}{2}\left\langle S\uparrow\left|p_{x}\right|X\uparrow\right\rangle \mp\frac{1}{2}\left\langle S\uparrow\left|p_{y}\right|Y\uparrow\right\rangle \,,
\label{eq:CBup-HHup}
\end{eqnarray}
which is non-zero only for $p_{-}$, 
\begin{eqnarray}
\left\langle \textrm{CB}\Downarrow\left|p_{\pm}\right|\textrm{HH}\Downarrow\right\rangle  & = & \left\langle S\downarrow\left|\frac{p_{x}\pm ip_{y}}{\sqrt{2}}\right|\frac{i}{\sqrt{2}}\left(X-iY\right)\downarrow\right\rangle 
  =  \frac{i}{2}\left\langle S\downarrow\left|p_{x}\right|X\downarrow\right\rangle \pm\frac{i}{2}\left\langle S\downarrow\left|p_{y}\right|Y\downarrow\right\rangle \,,
\label{eq:CBdw-HHdw}
\end{eqnarray}
which is non-zero only for $p_{+}$,
\begin{eqnarray}
\left\langle \textrm{CB}\Uparrow\left|p_{\pm}\right|\textrm{LH}\Downarrow\right\rangle  & = & \left\langle S\uparrow\left|\frac{p_{x}\pm ip_{y}}{\sqrt{2}}\right|\frac{1}{\sqrt{6}}\left[\left(X-iY\right)\uparrow+2Z\downarrow\right]\right\rangle 
 =  \frac{1}{2\sqrt{3}}\left\langle S\uparrow\left|p_{x}\right|X\uparrow\right\rangle \pm\frac{1}{2\sqrt{3}}\left\langle S\uparrow\left|p_{y}\right|Y\uparrow\right\rangle \,,
\label{eq:CBup-LHdw}
\end{eqnarray}
which is non-zero only for $p_{+}$, and 
\begin{eqnarray}
\left\langle \textrm{CB}\Downarrow\left|p_{\pm}\right|\textrm{LH}\Uparrow\right\rangle  & = & \left\langle S\downarrow\left|\frac{p_{x}\pm ip_{y}}{\sqrt{2}}\right|\frac{i}{\sqrt{6}}\left[\left(X+iY\right)\downarrow-2Z\uparrow\right]\right\rangle 
  =  \frac{1}{2\sqrt{3}}\left\langle S\downarrow\left|p_{x}\right|X\downarrow\right\rangle \mp\frac{1}{2\sqrt{3}}\left\langle S\downarrow\left|p_{y}\right|Y\downarrow\right\rangle \,,
\label{eq:CBdw-LHup}
\end{eqnarray}
which is non-zero only for $p_{-}$.
\end{widetext}

In addition to Eqs.~(\ref{eq:CBup-HHup})--(\ref{eq:CBdw-LHup}),  we can conclude 
that $\left\langle \textrm{CB}\Uparrow\left|p_{\pm}\right|\textrm{HH}\Downarrow\right\rangle =\left\langle \textrm{CB}\Downarrow\left|p_{\pm}\right|\textrm{HH}\Uparrow\right\rangle =0$ 
and $\left\langle \textrm{CB}\Uparrow\left|p_{\pm}\right|\textrm{LH}\Uparrow\right\rangle =\left\langle \textrm{CB}\Downarrow\left|p_{\pm}\right|\textrm{LH}\Downarrow\right\rangle =0$, independent of  the light polarization.

%===============================================================================

%===============================================================================


\begin{thebibliography}{99}
\bibitem{note_gain}
This gain coefficient corresponds to the ratio of the number of photons emitted 
per second per unit volume and the number of injected photons per second per unit 
area, therefore having a dimension of 1/length.
\bibitem{note_rate_holes}
Within the rate-equation description, including a widely-used spin-flip model, the
$n_+p_+$ recombination only gives $S^-$ helicity of the emitted light.
From the consideration this means that there is only one type of holes 
within  a four-band model (CB and VB with a two-fold spin degeneracy).
\bibitem{note_polarizations}
Analogous expressions can be introduced for the spin polarization of injection
and polarization of photon density.
\bibitem{note_nonmagnetic} 
We assume that there is no intrinsic magnetic character of the laser due to
an applied magnetic field or the presence of a magnetic region. %IZ no MCD?
\bibitem{note_phase} 
A simple phase condition for the standing wave-pattern [recall Fig.~2(a)] can be  written as $L=m\lambda/(2n)$,\cite{Michalzik:2013}
where $L$ is the cavity length, $m$ is the mode index, and $\lambda$ is the wavelength of the emitted light. 
The polarization-dependence of the refractive index, $n$, thus leads to the polarization-dependence of the emitted frequency.
\bibitem{note_dope}
We assume a typical description of an active region which is usually undoped and
has a charge neutrality, $n=p$.
\bibitem{note_harddrive}
Commercial magnetic hard drives already employ ferromagnets with perpendicular anisotropy
which provides storing a higher information density.
\end{thebibliography}
\end{document}